\documentclass{osa-article}
\usepackage{amsmath}

\DeclareMathOperator*{\argmin}{arg\,min}

\journal{oe}

\articletype{Research Article}
\usepackage{siunitx}

\begin{document}

\title{Machine Learning Resistant Amorphous Silicon Physically Unclonable Functions (PUFs)}

\author{Velat Kilic\authormark{1}, Neil Macfarlane\authormark{1}, Jasper Stroud\authormark{1}, Samuel Metais\authormark{1}, Milad Alemohammad\authormark{1}, A. Brinton Cooper\authormark{1}, Amy C. Foster\authormark{1}, Mark A. Foster\authormark{1,*}}

\address{\authormark{1}Department of Electrical and Computer Engineering, Johns Hopkins University, Baltimore, U.S.A}

\email{\authormark{*}mark.foster@jhu.edu} 



\begin{abstract}
We investigate usage of nonlinear wave chaotic amorphous silicon (a-Si) cavities as physically unclonable functions (PUF). Machine learning attacks on integrated electronic PUFs have been demonstrated to be very effective at modeling PUF behavior. Such attacks on integrated a-Si photonic PUFs are investigated through application of algorithms including linear regression, k-nearest neighbor, decision tree ensembles (random forests and gradient boosted trees), and deep neural networks (DNNs). We found that DNNs performed the best among all the algorithms studied but still failed to completely break the a-Si PUF security which we quantify through a private information metric. Furthermore, machine learning resistance of a-Si PUFs were found to be directly related to the strength of their nonlinear response.
\end{abstract}

\section{Introduction}
The pervasive introduction of interconnected devices into our lives has generated increased concerns for privacy and has made cryptographic systems more important than ever before. Many crypto protocols rely heavily on the security of keys that are stored in device memory and are susceptible to malware attacks. Physically unclonable functions (PUFs) have been proposed as an alternative \cite{Pappu} whose response to external stimuli (challenge) is determined by their microscopic structure, which is difficult to clone. PUF operation consists of two phases: enrollment and deployment. During the enrollment process, the manufacturer creates a challenge response pair (CRP) library by probing the device with unique binary challenges and measuring/generating the corresponding digitized response. The CRP data set is then stored for the deployment phase where the PUF device can be authenticated by probing it with a subset of challenges in the CRP data set and comparing the responses. To be a strong security primitive a PUF must exhibit behavior that is i) deterministic, ii) unpredictable, and iii) unique. Additionally, iv) the device should be unclonable in that its behavior cannot be replicated by a copy of the device. An unclonable but deterministic pseudo-random function generator makes a desirable PUF. 

Since their inception, PUFs built on various platforms have been proposed \cite{Zhang,Grubel,Geis,Mesaritakis}; however most, if not all, electronic PUFs have been shown to be susceptible to modelling and machine learning (ML) attacks \cite{Ruhrmair} and therefore do not satisfy one of the four requirements of a strong security primitive. In contrast, bulk optical scattering PUFs have remained resistant to machine learning attacks due to the sheer size of their challenge-response space \cite{Horstmeyer_host}. However, such bulk optical scattering PUFs are large, unreliable, and difficult to integrate into conventional electronics. ML resistance of scattering PUFs rely on high dimensional challenge-response space which is physically upper bounded by the number of optical modes and therefore device size. To solve this trade-off between security and practicality, a novel PUF architecture has been recently demonstrated which is known as a silicon photonic PUF \cite{Grubel,MacFarlane,Stroud,Atakhodjaev}. Silicon photonic PUFs are devices fabricated using conventional CMOS-compatible planar semiconductor fabrication but leverage nonlinear optical wave physics for their challenge-response behavior. Thus silicon photonic PUFs strike a balance between the practicality of electronic PUFs and the security offered by optical PUFs. To this end nonlinear crystalline silicon (c-Si) photonic PUFs have been recently shown to be resistant to modeling attacks based on deep neural networks (DNNs) and the degree of resistance was shown to be correlated with the strength of the optical nonlinearity \cite{Bosworth}.

In this work, we introduce PUFs based on hydrogenated amorphous silicon (a-Si:H) which has an effective nonlinearity that is an order of magnitude larger than c-Si \cite{asi_nonlinear}. Additionally, high-speed parametric amplification has been shown in a-Si:H waveguides, showing a better nonlinear figure of merit (defined as the ratio of the Kerr coefficient ($n_2$) to the two-photon absorption coefficient ($\beta_{TPA}$) time wavelength) than c-Si at C-band wavelengths \cite{asi_fom1,asi_fom2}. This is caused both by the higher Kerr coefficient of a-Si:H, but also the lower TPA from a-Si:H having a larger band gap energy. Therefore, a-Si:H photonic PUFs have the potential to be even more resistant to modelling attacks due to increased nonlinearity and better integration. Also, we further study the machine learning resistance of nonlinear a-Si:H photonic PUFs through the application of a suite of learning strategies including linear regression, k-nearest neighbor, decision trees, and DNNs. We find that the PUF’s resistance to machine learning attacks is tied to its degree of optical nonlinearity and quantify this resistance in terms of the amount of private information.

\begin{figure*}[h]
\begin{center}
 \includegraphics[width=\textwidth]{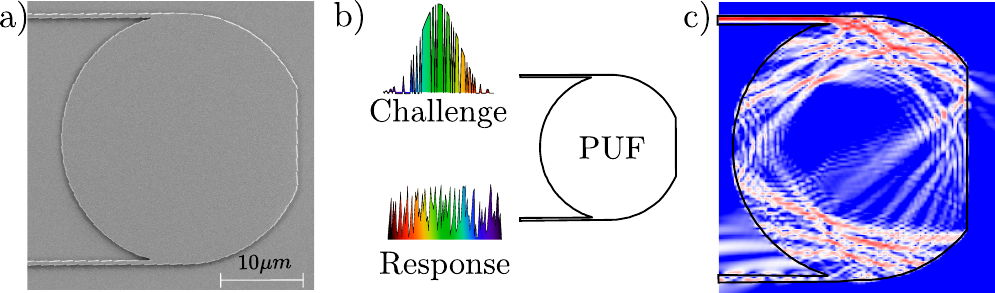}
 \caption{a) Scanning electron microscope (SEM) image of an amorphous silicon ray chaotic cavity used as a PUF. b) The device is probed by patterning a chirped pulse with the binary challenges and the output time series signal is measured in the Hadamard basis.  c) Finite difference time domain (FDTD) simulation of a mode in the PUF device.}  \label{fig_1}
\end{center}
\end{figure*}

\section{A Nonlinear Integrated a-Si:H Photonic PUF}
In contrast to bulky scattering optical PUFs, we have designed an integrated a-Si:H photonic PUF as shown in Fig. \ref{fig_1}a which has two optical single mode wave-guides and a cavity. Single mode waveguide ensures stable and repeatable coupling of light in (and out of) the device using telecommunications commercial grade fiber optics, which proves more practical than using a large scattering medium. The geometry of the cavity is a clipped disc, which is known to be ray-chaotic. The input optical signal reverberates around the cavity: the behaviour is then similar to the one of an optical scattering PUF, where reverberation is used to limit its size. The surface roughness of the boundaries, that comes from the fabrication variation help randomize the cavity behaviour, which makes cloning the PUF intrinsically challenging as that would require near atomic level precision in both measurement and fabrication. Furthermore, integrated photonic devices are very stable in time, with consistent behaviour over multiple days\cite{Grubel}. Yet, the limited size of the device limits its degrees of freedom (size of the independent pair-response challenges space), and so we make use of the large $\chi^{(3)}$ nonlinearity of a-Si:H \cite{Li} inside the chaotic cavity to add another layer of complexity on the intrinsic function of this object, which should further protect it against machine learning attacks.

Ray chaotic design is necessary to protect against nearest neighbor attacks. PUF response needs to decorrelate fast even for a small change in the challenge (i.e large Lyapunov exponent). Otherwise, a large measurement database can be used to find a similar enough challenge-response pair. Ray chaoticity along with Kerr nonlinearity provides protection against these attacks because a change in the challenge leads to a change in the intensity which modulates the cavity index through $\chi^{(3)}$. A ray chaotic cavity can then translate even small changes in refractive index to large variations in the cavity response.

\subsection{Fabrication}
The a-Si:H devices are fabricated using mature CMOS fabrication techniques. Fabrication begins with a 4-inch silicon wafer with 3000-nm of thermal oxide on the surface. First, 250-nm of a-Si:H is deposited using plasma-enhanced chemical vapor deposition (PECVD) at 350\textdegree C with 200-SCCM of silane gas, and 800 SCCM of argon gas. Then 150-nm of $\mathrm{SiO_2}$ is deposited also using PECVD to act as a hard mask for etching. The devices are patterned using electron-beam lithography (EBL). The devices are then etched using reactive ion etching (RIE) and inductively coupled plasma etching (ICP). Finally, about 1000-nm of $\mathrm{SiO_2}$ is deposited using PECVD to act as an upper cladding and optically isolate the devices from air.      

\begin{figure*}[h]
\begin{center}
 \includegraphics[width=\textwidth]{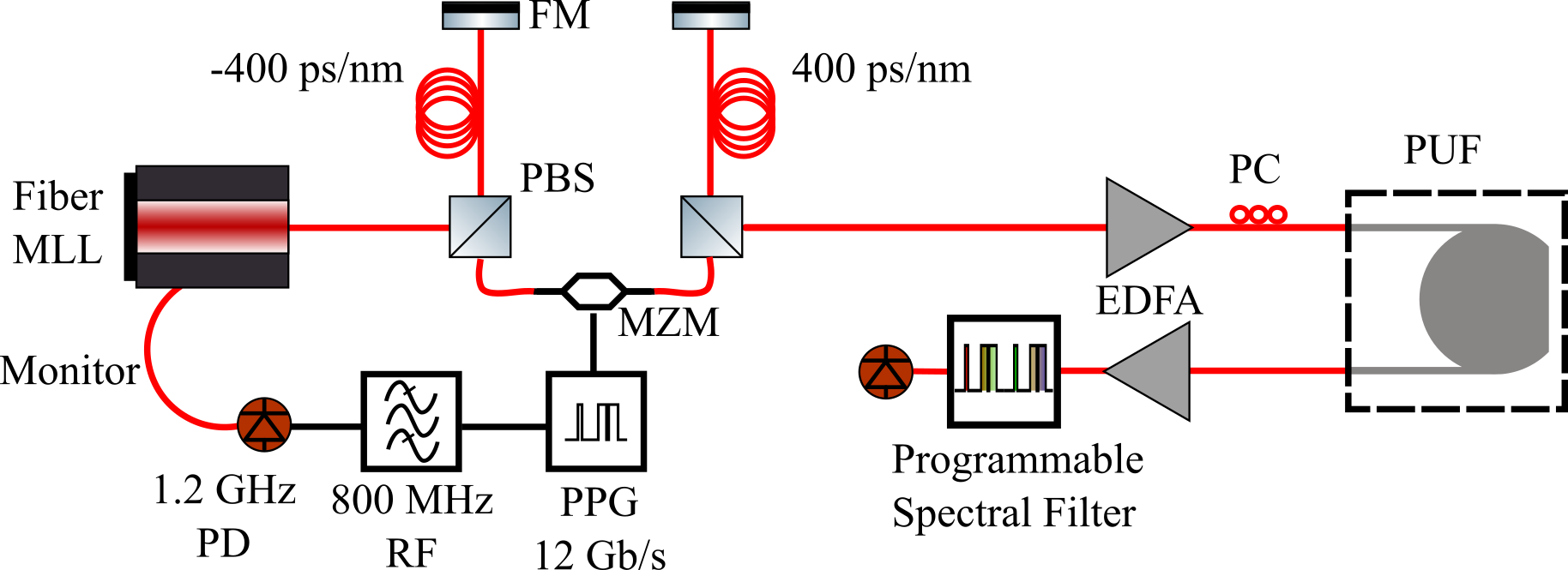}
 \caption{Ultrafast optical setup for creating the CRP (challenge-response pair) library. Pulses from the MLL (mode locked laser) are stretched using dispersive fiber. Pulses are then patterned by an MZM (Mach-Zehnder modulator) modulated by a PPG (pulse pattern generator) triggered by the 8th harmonic of the MLL. The pulses are then compressed with equal but opposite amounts of dispersion. They are then fed through an EDFA (Erbium doped fiber amplifier), PC (polarization controller), PUF (physically unclonable function), EDFA again, and finally a programmable spectral filter.}  \label{fig_sch}
\end{center}
\end{figure*}

\subsection{Experiment}
Authentication in PUF systems is often done by a process known as challenge-response verification. Upon manufacturing, each device is fully characterized by applying a sequence of binary input keys (i.e. challenges). The challenge-response library (CRL) is then constructed as a library of all unique, and unclonable responses to individual challenges. Authentication is carried out by comparing the PUF response to a known challenge, against the response stored in the library. 

In order to achieve an extremely large challenge-response space, we use a spectrally encoded ultra-fast optical scheme shown in Fig \ref{fig_sch}. The ultra-fast encoder module comprises of a dispersive fiber which performs the frequency-to-time mapping of mode-locked laser (MLL) pulses. Dispersed pulses enter a high speed electro-optic modulator, driven by a pulse pattern generator (PPG). The PPG is synchronized to the laser pulses and each pulse is encoded with a known sequence of 336 pseudo-random bits (i.e. challenges). After patterning, the laser pulses are temporally compressed using a complimentary dispersion compensating fiber. Compressed pulses are further amplified using an Erbium Doped Fiber Amplifier (EDFA) in order to increase the peak power required for non-linear optical interactions inside the silicon PUF device. Inside the PUF device, each encoded laser pulse experiences a unique, complex spectro-temporal modulation due to a combination of linear and nonlinear optical interactions. It is important to note that these interactions will vary based on the choice of the binary challenge sequence. These pulses are then detected using a photodetector and the pulse peaks are digitized with 14 bit resolution. By using the orthogonal Hadamard spectral filter at the output of the PUF device we effectively extract 127$\times$14 bits of key data for each binary challenge sequence.

We use fractional Hamming distance (FHD) to compare the responses from two challenges
\begin{equation}
    FHD(a,b) = \frac{1}{L}\sum_{i=1}^L a_i \oplus b_i
\end{equation}
where $a$ and $b$ are bit strings (i.e. response sequence) of length $L$ and $\oplus$ is the xor operation. Calculating the FHD results in values between $0$ (similar) and $0.5$ (dissimilar). Ideally, true PUF responses would achieve an FHD of $0$ and the clone would achieve $0.5$ with very small variance, although in practice, this is rarely the case due to environmental factors (i.e temperature, humidity, etc.) and noise.

In the following section, we investigate the machine learning resistance of our nonlinear silicon photonic PUF device. Modelling attacks are performed on the the Hadamard transform of the analog time series signal. In other words, we assume snooping of an overt channel (in an application, the PUF security can be further strengthened by techniques such as histogram equalization). For comparison however digitized responses are used. 

\begin{figure*}[!h]
\begin{center}
 \includegraphics[width=\textwidth]{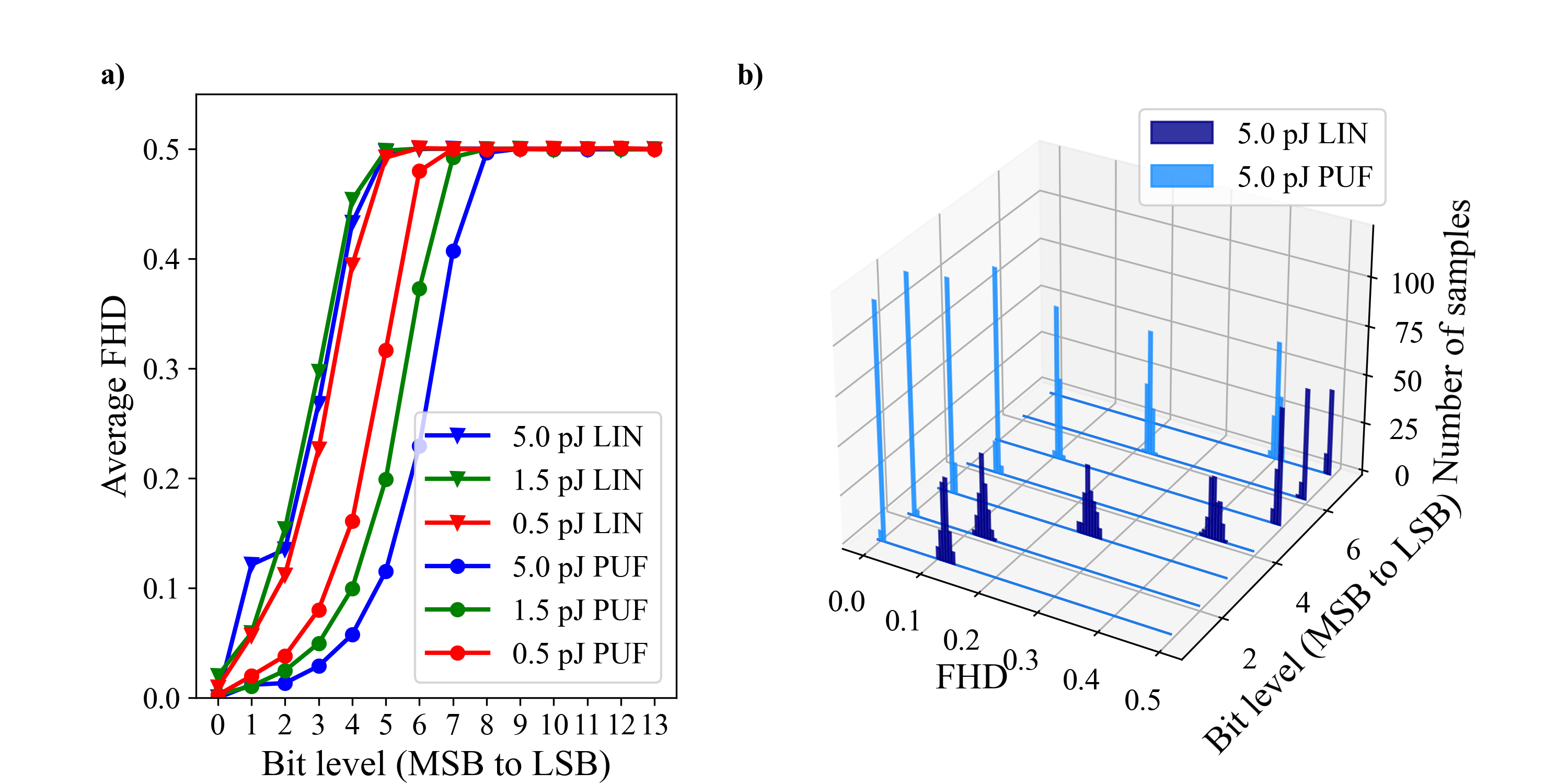}
 \caption{a) Average fractional Hamming distance (FHD) between PUF responses as well as linear regression guess as a function of the bit level (from the most significant bit to the least significant bit). Different colors correspond to different pulse energies (blue: 5pJ, green: 1.5 pJ, red: 0.5 pJ). Circles correspond to FHD between PUF responses probed twice for the same challenges and is a measure of the repeatability of the measurement. Triangles correspond to FHD between ground truth PUF response and the linear regression guess. LIN stands for linear model. b) Histogram of FHDs for the 5 pJ pulse energy for a 20kbit key length both for the PUF and the linear model prediction.}  \label{fig_2}
\end{center}
\end{figure*}

\begin{figure}[!h]
\begin{center}
 \includegraphics[width=\textwidth]{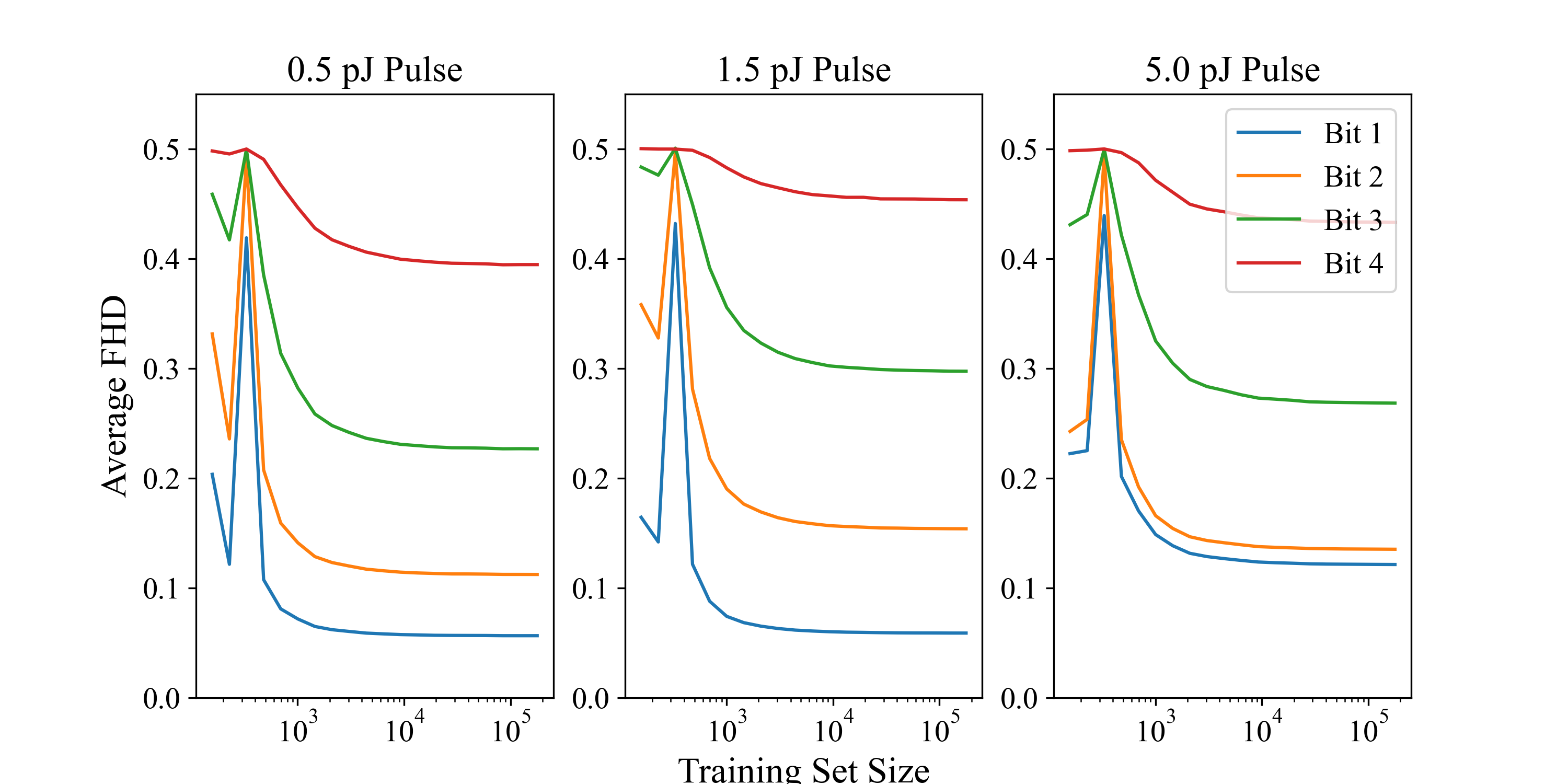}
 \caption{Average fractional Hamming distance as a function of training data set size for bit levels 1-4. Training set size beyond a few thousand samples provides diminishing returns.}  \label{fig_tr_size}
 \end{center}
\end{figure}

\section{Machine Learning Attacks}
\subsection{Linear Regression}
Optical response of the PUF can be expanded into linear and nonlinear parts which are large and perturbative respectively. Linear component is easy to learn by using linear regression and the number of measurements needed is small (see Section 4). Here we investigate how much of the PUF response is linear and can be learned by solving the following linear regression problem:
\begin{equation}
    \hat{W}_{lin} = \argmin_W \lvert W \mathbf{x} - \mathbf{y} \rvert^2_2 
\end{equation}
where $\mathbf{x}$ and $\mathbf{y}$ are data matrices of the challenges and the corresponding responses. A vector of ones is appended to the challenges to compute the bias term. Training and test sets consist of 180,000 and 20,000 challenge response pairs respectively. Optimization problem is solved by using \textit{Scikit-learn}, a machine learning toolbox\cite{sklearn}. Training is performed on the analog information and the results are then digitized (14 bits). Finally, the fractional Hamming distance (FHD) between the prediction and the ground truth is calculated. Ground truth is generated by probing the PUF twice with the same challenge and averaging the responses. Fig. \ref{fig_2}a shows the results for the linear regression for the same device at three different pulse energy levels (0.5, 1.5 and 5 pJ). As expected, the most significant bit which corresponds to roughly guessing the total power level and the linear contribution can be learned very well. However, following bits can be readily distinguished and as shown in \ref{fig_2}b, a decision boundary can be very clearly drawn. This supports our claim that integrated PUF security is derived from the perturbative nonlinear response.

Ideally, FHD between PUF responses would be very small and the FHD between PUF responses and the prediction would be 0.5 corresponding to a random guess. As shown in Fig. \ref{fig_2}, increasing the pulse energy increases the gap between the prediction and the ground truth value which is achieved through two mechanisms in this case: i) increased pulse energy leads to more nonlinear interactions and ii) signal to noise ratio increases which makes the responses more repeatable. Therefore, the largest gap is seen between the blue curves(higher energy) and the smallest between the red curves(lower energy).

Although, the linear response can be learned with the given dataset, an attacker often will not have access to such a large CRP data set. In order to test the effect of data set size, we randomly sampled a fraction of the set to train a linear model. As shown in Fig. \ref{fig_tr_size}, increased CRP set size leads to better prediction but with diminishing returns after a few thousand samples. This number is determined by the number of modes in the optical device, receiver/transmitter degrees of freedom as well as SNR which we investigate in section 4.

\subsection{k-Nearest Neighbor (kNN)}
k-Nearest neighbor is a \textit{lazy learning} algorithm that keeps all the training data without any generalization. Test data is compared to all the samples in the training set and the output of the k-closest samples are used in making the final prediction. Since kNN does not require any modelling, it might be suitable for predicting highly nonlinear data given a large training set.

\begin{figure}[!h]
\begin{center}
 \includegraphics[width=\textwidth]{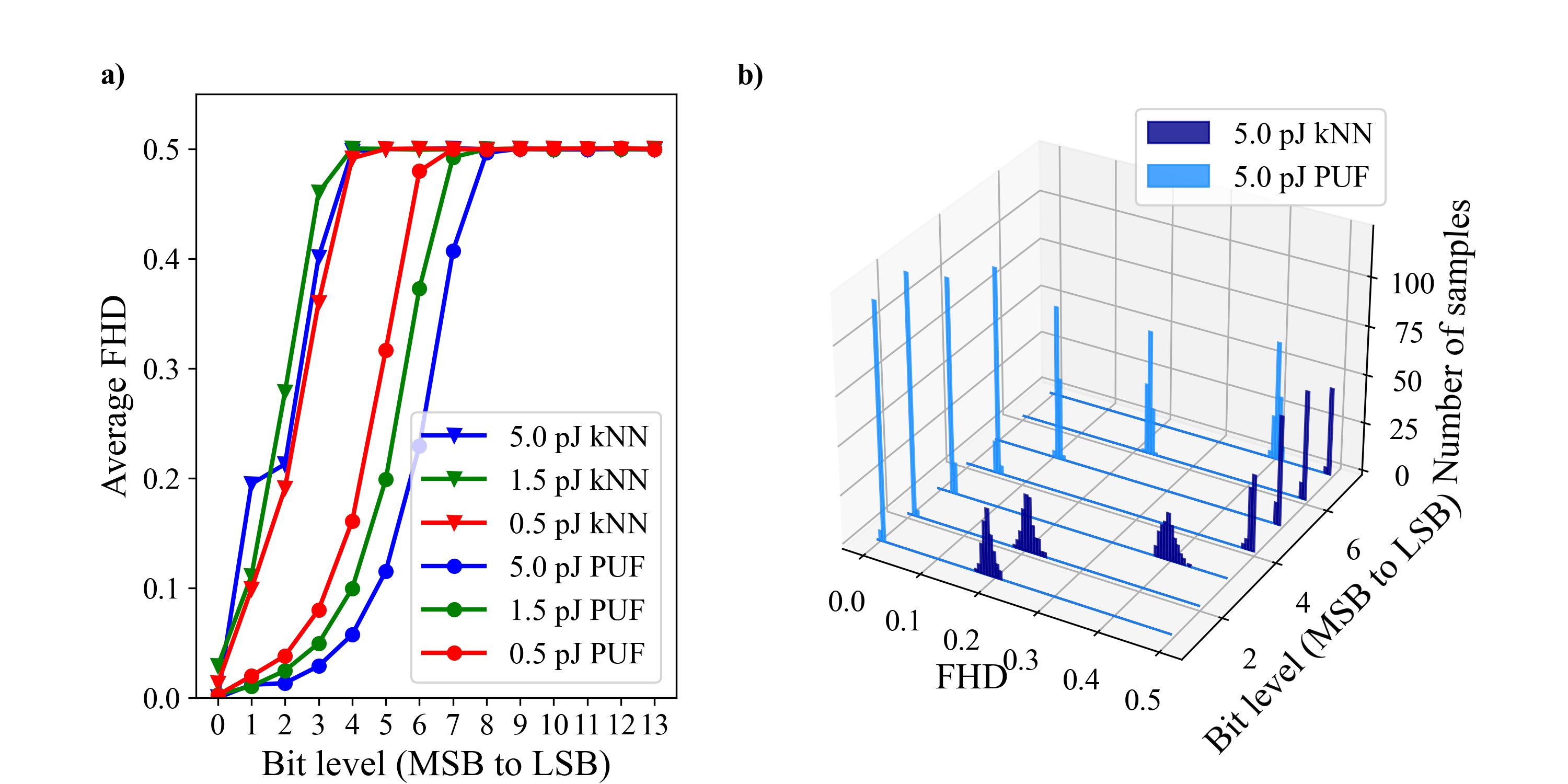}
 \caption{Average FHD at 3 different power levels as a function of the bit level. Triangles correspond to kNN predictions and circles correspond to PUF ground truth (displayed for comparison). b) FHD histogram at each bit level for a key length of 20kbits.}  \label{fig_kNN}
 \end{center}
\end{figure}

However, as shown in Fig. \ref{fig_kNN} and \ref{fig_All_Models}, kNN algorithm performed worse than the linear regression model. The training set had 180k samples and the test set had 20k samples. \textit{Scikit-learn} implementation is used with Euclidean distance as the similarity metric and the best performance was achieved for $k=100$ \cite{sklearn}. Performance of the kNN algorithm in general depends on the neighbors exhibiting similar behavior and we observe that it performs the worst for data corresponding to high energy pulses which shows nonlinear response is key to ML resistance. Physically, as the pulse energy is varied, the refractive index profile of the PUF changes due to Kerr nonlinearity. Ray chaotic cavity then maps small changes of the refractive index to large changes in device response.

Due to non-flat laser pulse profile, not all bits in the challenge vectors have the same importance. In fact, matrix weights from the linear regression model obtain larger values for the challenge bits near the pulse energy maximum. Therefore, we implemented a hybrid linear regression - kNN algorithm that assigns larger weights to these bits in the challenge vectors. However, the hybrid model still performs worse than a simple linear regression.

\begin{figure}[!h]
\begin{center}
 \includegraphics[width=\textwidth]{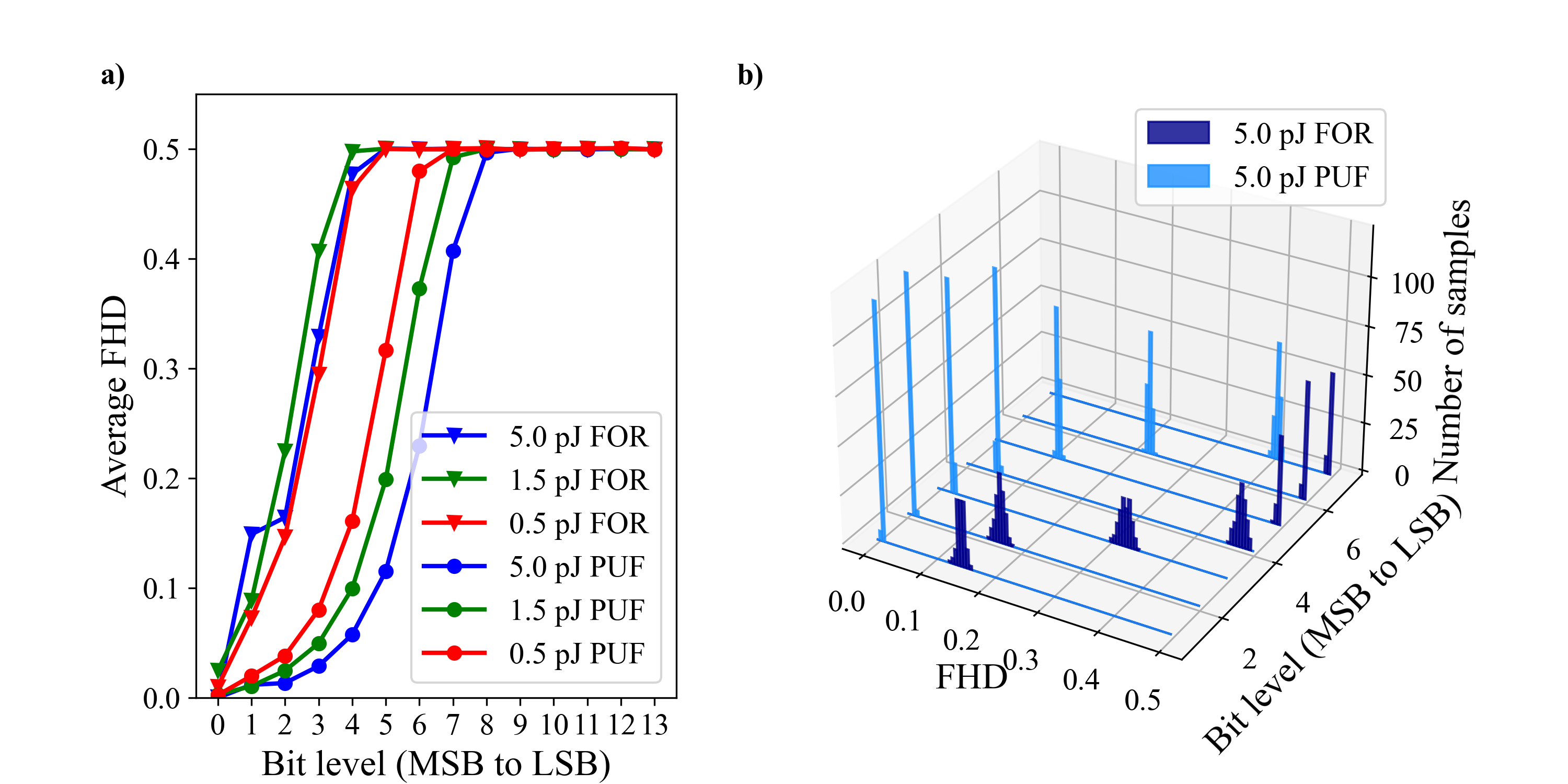}
 \caption{Average FHD at 3 different power levels as a function of the bit level. Triangles correspond to random forest (FOR) predictions. b) FHD histogram at each bit level for a key length of 20kbits.}  \label{fig_Ran_For}
 \end{center}
\end{figure}

\begin{figure}[!h]
\begin{center}
 \includegraphics[width=\textwidth]{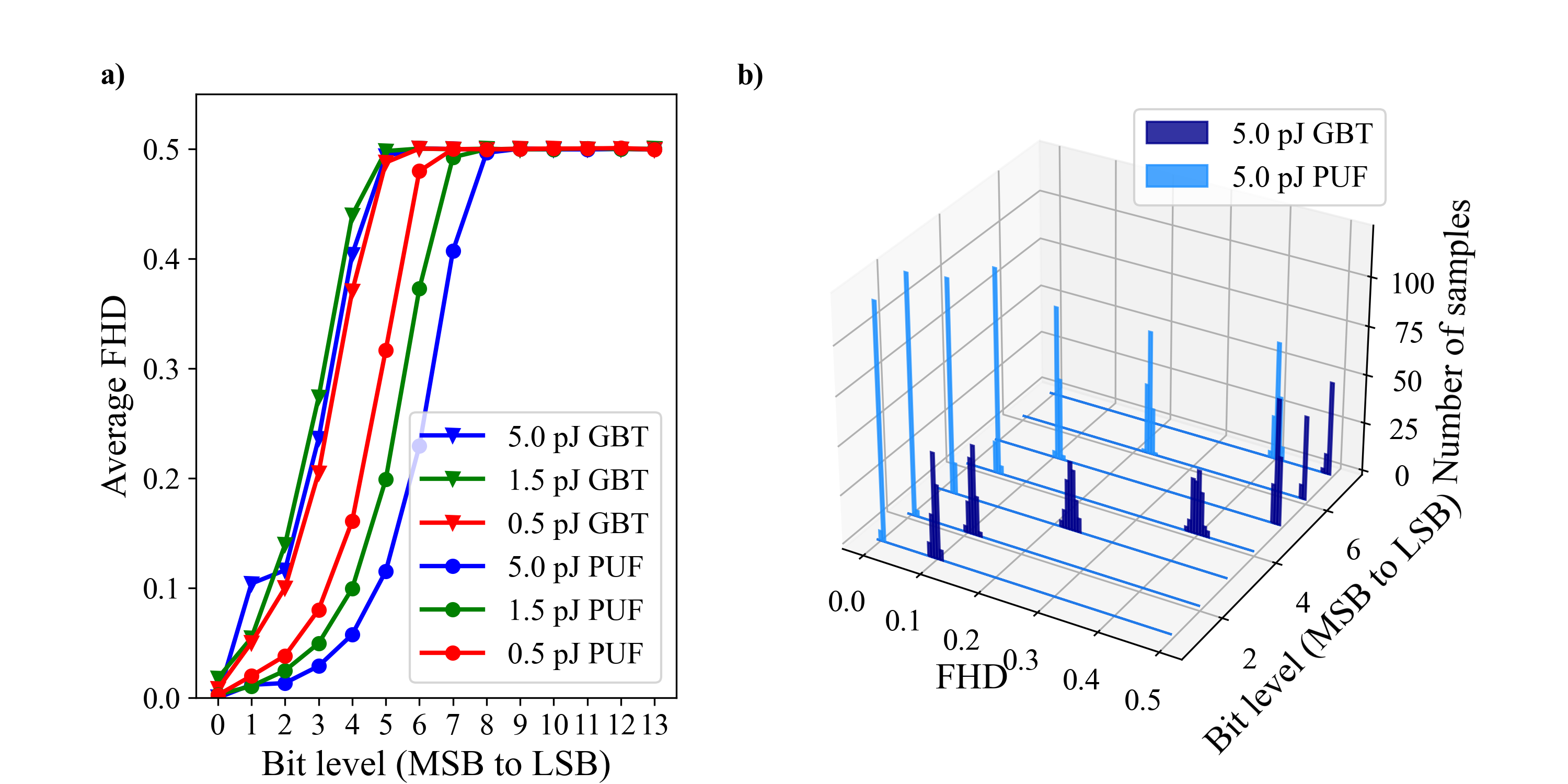}
 \caption{Average FHD at 3 different power levels as a function of the bit level. Triangles correspond to gradient boosted tree (GBT) predictions. b) FHD histogram at each bit level for a key length of 20kbits.}  \label{fig_Decision_Tree}
 \end{center}
\end{figure}

\subsection{Decision Tree Models}
Decision trees combine multiple levels of \textit{if-else} statements to make a prediction. Decision trees are often used in ensembles such as random forests since they are prone to over-fitting. Getting a consensus vote from multiple weak predictors improves accuracy and reduces variance. We used \textit{Scikit-learn} implementation of random forests with 100 decision trees per forest. Results from this algorithm are shown in Fig. \ref{fig_Ran_For} which performed worse than linear regression. We also used gradient boosting which constructs trees one at a time to compensate weakness of the existing trees in the ensemble as opposed to random forests which randomly constructs them. For that purpose, we use a state of the art algorithm \textit{XGBoost} \cite{xgboost} which performed significantly better as shown in \ref{fig_Decision_Tree}; better than linear regression but worse than DNNs as shown in Fig. \ref{fig_All_Models}. Implementation details of XGBoost are beyond the scope of this work.

\subsection{Deep Neural Networks (DNNs)}
\begin{figure}[!h]
\begin{center}
 \includegraphics[width=\textwidth]{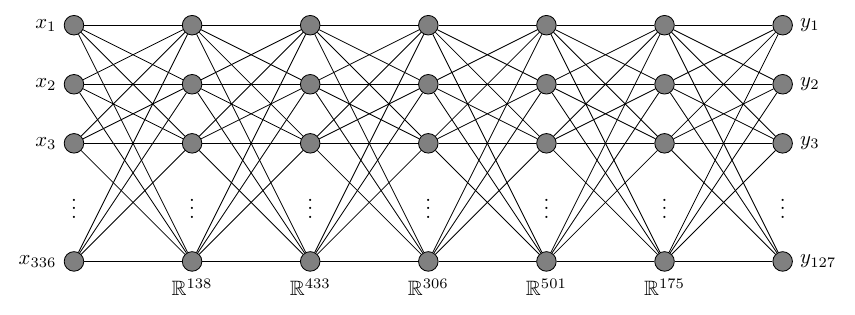}
 \caption{Deep neural network with rectified linear unit (ReLU) activation function in the hidden layers. Input is a 336 bit long challenge and the output is an analog signal of length 127.}  \label{fig_dnn_net}
 \end{center}
\end{figure}

With the availability of powerful GPUs, deep neural networks have gained popularity in the past years and have outperformed other machine learning algorithms across many tasks from computer vision to natural language processing and decision making \cite{LeCun2015}. DNNs can be mathematically described as linear transformations followed by nonlinearities where parameters of the network can be optimized using gradient descent. In this work, we use Pytorch, an open source software platform for building neural networks to implement the DNN shown in Fig. \ref{fig_dnn_net}. All the layers except for the output use the rectified linear unit (ReLU) as the nonlinear activation function. Training, validation and test sets consisted of 179k, 1k and 20k samples respectively. The validation set is used to optimize the hyperparameters of the network such as number of hidden layers and their size, optimizer type, and learning rate using a hyperparameter optimizer Optuna \cite{optuna}. A batch size of 200 is used and training is performed for 200 epochs. Adam is used as the optimizer \cite{adam} with a learning rate of $4.99\times10^{-4}$.

\begin{figure}[!h]
\begin{center}
 \includegraphics[width=\textwidth]{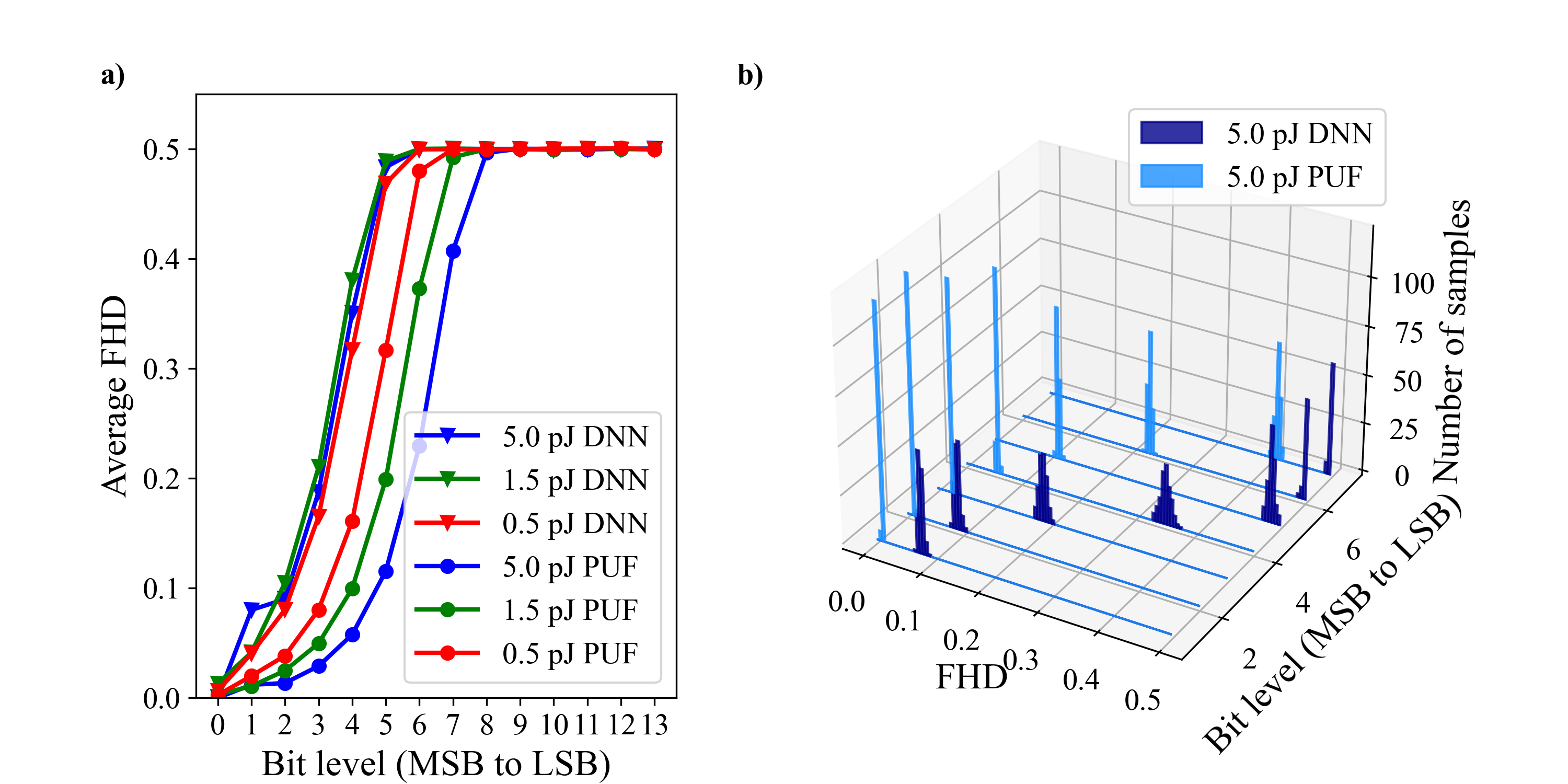}
 \caption{Average FHD at 3 different power levels as a function of the bit level. Triangles correspond to the deep neural network (DNN) predictions. b) FHD histogram at each bit level for a key length of 20kbits.}  \label{fig_dnn}
 \end{center}
\end{figure}

\begin{figure}[!h]
\begin{center}
 \includegraphics[width=\textwidth]{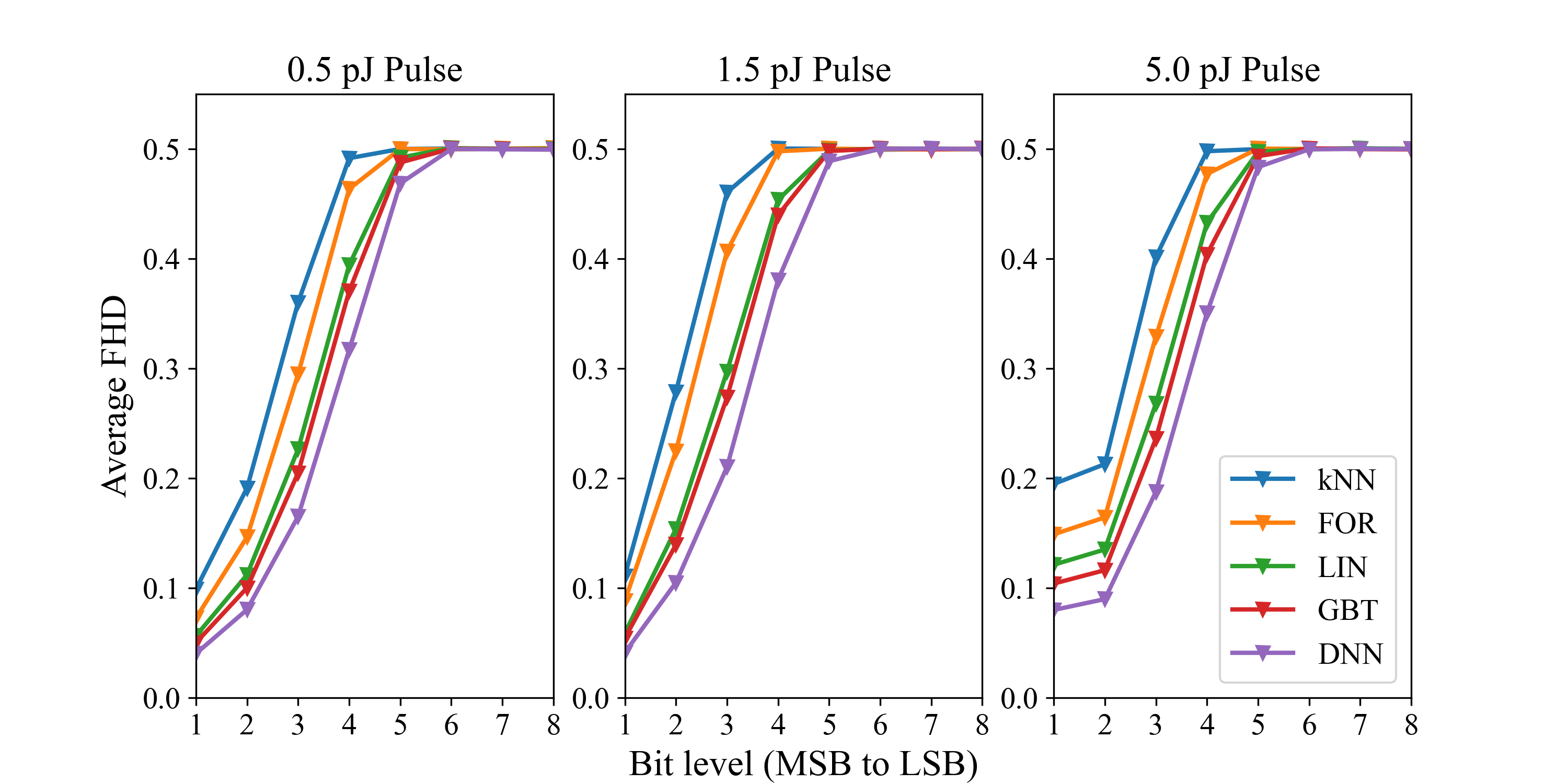}
 \caption{Average FHD comparison for all models at different pulse energies a) 0.5 pJ, b) 1.5 pJ, c) 5.0 pJ. ML attack performance remains the same across different pulses energies with DNN performing the best and kNN the worst.}  \label{fig_All_Models}
 \end{center}
\end{figure}

As shown in Fig. \ref{fig_dnn} and \ref{fig_All_Models}, DNN provided the best predictions. However, as seen in Fig. \ref{fig_dnn}b a decision boundary can still be drawn accurately. Indeed, this improvement comes from the ability of the neural network to model some of the device non-linearity. In order to better study our PUF's security, we investigate its private information content in the following section.

\subsection{Private Information}
PUFs can be used to securely transmit $H$ bits of information as determined by entropy of the FHD distribution. In the presence of an attack, some of that information is revealed through modelling and thus private information is reduced. Therefore, we introduce a private information metric which is a measure of the amount of information that can be transmitted securely in the presence of an attack. We define private information as a symmetric, averaged Kullback–Leibler (KL) divergence which is also known as Jeffrey divergence:
\begin{equation}
    PV(p,q) = \frac{1}{2} \left( KL(p,q) + KL(q,p) \right)
\end{equation}
where
\begin{equation}
    KL(p,q) = -\int^{\infty}_{-\infty} p\left(x\right) ln\left( \frac{q(x)}{p(x)} \right)
\end{equation}
$p(x)$ and $q(x)$ are probability density functions (in this case that of PUF and model FHD distributions for 20kbit keys). We assume FHD histograms are generated through a Gaussian process so the private information metric can be further simplified:
\begin{equation} \label{eq:pv}
    PV(p,q) = \frac{1}{2} \frac{\left( \sigma_1^2 - \sigma_2^2 \right)^2 + \left( \mu_1 - \mu_2 \right)^2 (\sigma_1^2 + \sigma_2^2)}{2\sigma_1^2\sigma_2^2}
\end{equation}
where $\sigma_{1,2}$ and $\mu_{1,2}$ are the standard deviations and means of $p(x)$ and $q(x)$ respectively. Private information in this form is in natural unit of information (nats) but can be converted to bits: 1 bit = $\frac{1}{ln(2)}$ nat. In the limit $\sigma_1 \approx \sigma_2 \approx \sigma$, eq. \ref{eq:pv} simplifies to $\frac{\Delta\mu^2}{2\sigma^2}$ which makes intuitive sense since private information is maximal for narrow distributions and large separation of the means between the model and the real PUF histograms.

\begin{figure}[!h]
\begin{center}
 \includegraphics[width=0.8\textwidth]{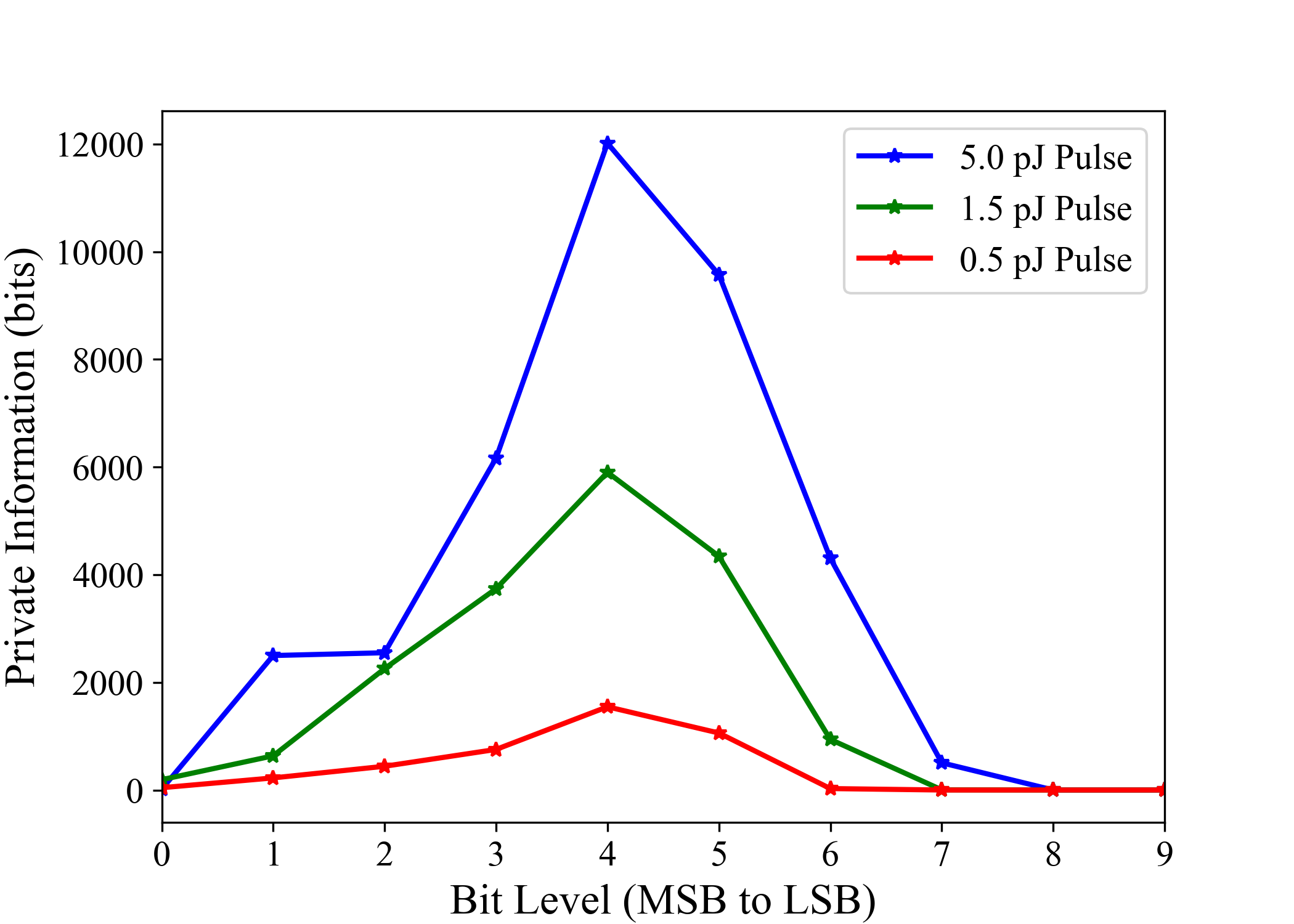}
 \caption{Private information measured in bits as a function of the bit level for different pulse energies in the presence of a DNN attack for 20kbit keys. Bit level 4 retains the largest amount of private information striking a balance between ML resistance and SNR trade-off.}  \label{fig_priv}
 \end{center}
\end{figure}

In Fig. \ref{fig_priv}, we plot private information in the presence of a DNN attack as a function of the bit level for different pulse energies. Although, the same plot can be generated for all other attacks, DNN is picked since it performs the best as shown in \ref{fig_All_Models} and sets an upper limit on the private information. As expected, total private information increases as the pulse energy increases due to increased nonlinearity. Further, it is observed that the bit levels 2-7 have the largest amount of private information as they balance enhanced security from nonlinear effects at the corresponding precision level and random noise.

\section{Linear Collapse Problem for PUFs Based on Linear Optical Devices}
Traditional electronic PUFs scale in complexity through the addition of discrete unit cells of circuit elements. In contrast linear optical PUFs leverage optical wave physics which scale in complexity with the number of optical modes. Small private information content of the first few bit levels (as shown in Fig. \ref{fig_priv}) in our a-Si PUFs is due to predictability of the linear contribution to the response. In this section, we provide theoretical bounds on PUFs based on linear optical devices and show how nonlinearity can improve device security.

It has been noted in the literature that large optical scattering PUFs contain many degrees of freedom, which implies a large CRP set \cite{Pappu}; however, this is not always true in practice due to limited degrees of freedom in the source and the receiver, which we call “linear collapse.” Even though the number of modes in the linear optical device might be very large, the degrees of freedom provided by the transmitter and/or the receiver significantly reduces the number of measurements needed to model the PUF for the purposes of the transmitter and the receiver. Let us define basis functions (modes) for each component: $\{ \lvert t_i \rangle \}$, $\{ \lvert c_j \rangle \}$, and $\{ \lvert r_k \rangle \}$ for the transmitter, the channel and the receiver respectively. Transmitter basis functions $\{ \lvert t_i \rangle \}$ encode the challenge information (i.e digital micromirror device (DMD) \cite{Horstmeyer_host}). Light emanating from the transmitter then interacts with the linear optical device (i.e scatterer) whose modes are $\{ \lvert c_j \rangle \}$ and is finally detected by the receiver (i.e camera). Suppose the transmitter has $P_1$ degrees of freedom (i.e at most the number of pixels of a DMD), the optical device has $M$ degrees of freedom (i.e number of modes) and the receiver has $ P_2$ degrees of freedom (i.e at most the number of pixels of a camera). Let us define the following operators $\hat{T}$, $\hat{C}$, and $\hat{R}$:
\begin{align}
\hat{T}_{M*P_1}&=\sum^{P_1}_{i=1} \sum^{M}_{j=1} T_{ij} \lvert c_j \rangle \langle t_i \rvert \\
\hat{C}_{M*M}&=\sum^{M}_{i=1} \sum^{M}_{j=1} C_{ij} \lvert c_j \rangle \langle c_i \rvert \\
\hat{R}_{P_2*M}&=\sum^{P_2}_{i=1} \sum^{M}_{j=1} R_{ij} \lvert r_j \rangle \langle c_i \rvert
\end{align}
where $\hat{T}$ maps transmitter modes to the channel modes, $\hat{C}$ maps the channel modes at the input to the channel modes at the output, and $\hat{R}$ maps the output channel modes to the receiver modes. Final mapping from transmitter to the receiver is then $\hat{D}_{P_2*P_1}~=~\hat{R}_{P_2*M}\hat{C}_{M*M}\hat{T}_{M*P_1}$. Usually $P_1 < P_2$ since PUFs are very similar to stream ciphers.

The PUF response for the purposes of the transmitter and the receiver can be modeled using $P_1$ independent measurements (i.e camera shots) even for a device with large number of modes. Of course, noise would increase the number of required measurements, whereas prior information on any part of the system (i.e sparsity, correlated pixels) would significantly reduce that number. In fact, optical scattering PUFs are being broken on a regular basis in computational imaging, and often with far fewer measurements than required by the Nyquist limit using compressed sensing \cite{Bertolotti,Shin_16,Shin_17,Antipa}. 

The end result is that linear optical scattering PUFs must leverage large scattering volumes and large and complex actuators and sensors with abundant degrees of freedom to achieve sufficient behavioral complexity to ensure a resistance to machine learning attacks. While this is possible in bulk geometries, this requirement prevents the application of linear optical scattering physics for integrated photonic geometries. Thus in integrated photonic geometries nonlinearity is crucial.

\section{Conclusion}
In conclusion, we have presented a CMOS compatible amorphous silicon photonic PUF architecture that uses nonlinear wave optics to achieve complex behavior. Complex behavior of the device can be modified by changing probe pulse energy. A large challenge response pair data set is used to implement machine learning attacks through linear regression, k-nearest neighbor, boosted decision trees, and deep neural networks (DNNs). Private information after a DNN attack is quantified since it provides an upper bound. DNN achieved the best performance however failed to break the PUF completely leaving as much as $8000$ bits of private information per bit level for $20$ kbit key length. Lastly, theoretical bounds for PUFs based on linear optical devices are presented. Future study will focus on a deeper level of integration with on-chip laser source and challenge generation as well as optical to digital read-out. Yet, even in its current form, compatibility of the PUF with telecommunications devices holds potential for secure communications.

\bibliography{sample}

\end{document}